\pdfoutput=1
\documentclass[aps,prl,preprint,onecolumn,amsmath,amssymb,groupedaddress]{revtex4-2}
\usepackage{graphicx}
\usepackage{chemformula} 
\usepackage{bm}
\usepackage{physics}
\usepackage{upgreek}
\usepackage{titlesec}

\titleformat{\section}[hang]{\normalfont\large\bfseries}{\thesection}{1em}{}
\titleformat{\subsection}[hang]{\normalfont\normalsize\bfseries}{\thesubsection}{1em}{}
\titleformat{\subsubsection}[hang]{\normalfont\normalsize\itshape}{\thesubsubsection}{1em}{}

\begin{document}

\title{Unidirectional chiral scattering from single enantiomeric plasmonic nanoparticles
}

\author{Yuanyang Xie*$^{,\#}$}
\author{Alexey V. Krasavin$^\#$}
\author{Diane J. Roth}
\author{Anatoly V. Zayats*}

\affiliation{Department of Physics and London Centre for Nanotechnology, King's College London, London, WS2R 2LS, UK \\
*Email Address: yuanyang.xie@kcl.ac.uk, a.zayats@kcl.ac.uk, $^\#$These authors contributed equally to this work.}

\date{\today}
\clearpage
\begin{abstract}
Controlling scattering and routing of chiral light at the nanoscale is important for optical information processing and imaging, quantum technologies as well as optical manipulation. Here, we introduce a concept of rotating chiral dipoles in order to achieve unidirectional chiral scattering. Implementing this concept by engineering multipole excitations in plasmonic helicoidal nanoparticles, we experimentally demonstrate enantio-sensitive and highly-directional forward scattering of circularly polarised light. The intensity of this highly-directional scattering is defined by the mutual relation between the handedness of the incident light and the chirality of the structure. The concept of rotating chiral dipoles opens up numerous possibilities for engineering of scattering from chiral nanostructures and optical nano-antennas for the design and application of chiral light-matter interaction.

\end{abstract}
 
\maketitle
\clearpage

Chirality describes a geometrical property of an object which cannot be superimposed with its mirror image using any combination of translation and rotation operations. Such geometrical characteristics, translated into enantio-sensitive optical and chemical properties, are of great importance in the physics of light-matter interaction, chiral chemistry and catalysis, as well as various areas of bioscience \cite{goerlitzer2021beginner,ke2023vacuum,luo2023large}. One of the most well-studied representations of material chirality in optics are circular birefringence and circular dichroism, resulting from different real and imaginary parts of the refractive index of the material, respectively, for circularly polarised light of opposite handedness \cite{wang2017circular,bliokh2016spin}. 

Phenomenologically, on a microscopic level, chiral optical properties are related to the excitation of collinear electric and magnetic dipoles, while the phase between them determines the handedness of the emitted or scattered light. In the case of rotating electric, Huygens and Janus dipolar sources \cite{eismann2018exciting, picardi2022integrated,liu2018generalized}, the interplay between orthogonal electric/electric, magnetic/magnetic and electric/magnetic dipoles through the near-field interference was extensively demonstrated for engineering counterintuitive unidirectional coupling to plasmonic and dielectric waveguides \cite{liu2012broadband} and far-field directional scattering \cite{elancheliyan2020tailored}, resulting in numerous applications in nanophotonics, metasurfaces, quantum technologies and metrology \cite{kuznetsov2016optically,picardi2019experimental,ballantine2020optical}. By engineering the amplitude and phase relation between the excited dipoles of plasmonic or dielectric nanoantennas, it is possible to control the directionality and spectral profile of the scattered light or achieve exotic nonradiative modes \cite{kaelberer2010toroidal,meng2022measuring,miroshnichenko2015nonradiating}. In contrast, for chiral dipolar sources, in the radiation patterns of the involved collinear magnetic and electric dipoles, the respective electric fields are orthogonal to each other and do not interfere, making it impossible to exploit the near-field interference to achieve directional properties.  

In this work, we develop a concept of rotating chiral dipoles to achieve highly-directional and enantio-sensitive scattering into purely circularly polarised light states, with the directionality controlled by the spin of the dipoles. We experimentally demonstrate this effect by engineering the phase-resolved excitation of electric and magnetic dipoles in chiral plasmonic nanohelicoids using a single-particle Fourier microscopy. 

The observed enantiomer-sensitive behaviour, which depends on the relation between the chirality sign of the nanoparticle and the handedness of the incident light, is important in designing and controlling polarisation-sensitive nanophotonics, enantiomeric optical forces, detection and sorting of chiral objects, as well as chiral photocatalysis.

\section{Results}
\subsection{Concept of rotating chiral dipoles} 
\begin{figure}[!ht]
    \begin{center}
        \includegraphics[width=9.9cm]{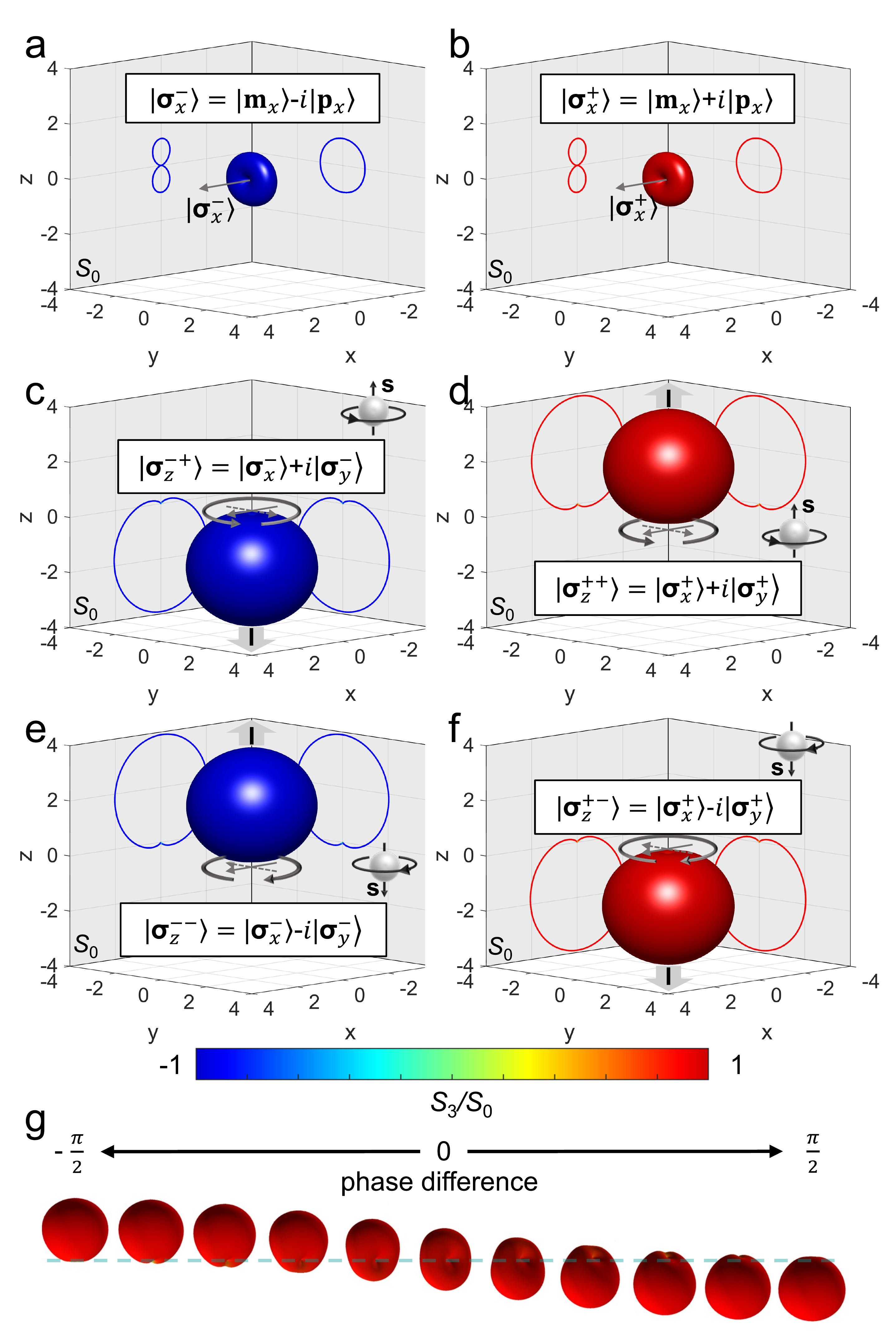}
        \caption{\textbf{Radiation patterns of chiral dipoles.} \textbf{a},~\textbf{b} Far-field radiation intensity and polarisation diagrams of (\textbf{a})~left-handed $\ket{\bm\upsigma^-_x}$ and (\textbf{b})~right-handed $\ket{\bm\upsigma^+_x}$ chiral dipoles, emitting pure RCP and LCP light, respectively. The arrows show the direction of the chiral dipole. \textbf{c}--\textbf{f}~Far-field radiation intensity diagrams of the rotating $\ket{\bm\upsigma^{\pm\pm}_z}$ chiral dipoles with indicated handednesses and rotation directions. The colour map shows $S_3/S_0$ Stokes parameter which represents the handedness of the emitted light (values 1 and -1 correspond to RCP and LCP, respectively, defined from the point of view of the source). The arrows show the rotation direction of the chiral dipole. \textbf{g}~Far-field radiation intensity andpolarisation diagrams of right-handed 'elliptically' polarised rotating chiral dipoles with varied phase difference between equal $x$- and $y$- components. The phase step is $\pi/10$.}\label{rotatingchiral}
    \end{center}
\end{figure}

In a direct analogy to electric and magnetic dipoles, chiral $\bm\upsigma^\pm$-dipoles emitting light with pure circular polarisation can be introduced, where '$+$' and '$-$' correspond to right circular polarised (RCP) and left circularly polarised (LCP) emission, respectively \cite{eismann2018exciting}. They can be represented within a dipolar framework as a sum of co-directed electric $\ket{\mathbf{p}}$ 
and a magnetic $\ket{\mathbf{m}}$ dipoles with a certain ratio of the magnitudes $p=m/c$ and a $\mp \pi/2$ phase shift (Fig.~\ref{rotatingchiral}a,~b) \cite{eismann2018exciting}.
Right- $\ket{\bm\upsigma^+_x}$ 
and left- $\ket{\bm\upsigma^-_x}$ handed chiral dipoles directed along the $x$-axis are given by $\ket{\bm\upsigma^{\pm}_x} = \ket{\mathbf{m}_x} \pm i \ket{\mathbf{p}_x} = (1,0,0) \ket{\bm\upsigma^{\pm}} = (1,0,0)\ket{\mathbf{m}} \pm i (1,0,0)\ket{\mathbf{p}}$. 
Chiral dipoles radiate purely circularly polarised light in a direction normal to the dipole orientation with a typical dipolar radiation pattern (Fig.~\ref{rotatingchiral}a,b). In contrast, a rotating electric dipole radiates circularly polarised light of opposite handedness in the opposite directions (Supplementary Fig.~1a). 

 In a direct analogy to rotating electric dipoles, two perpendicular chiral dipoles with a $\pm \pi/2$ phase shift generate a rotating chiral dipole (see Supplementary Fig.~6 for details). Right-handed ($+$) chiral dipole rotating clockwise ($+$) or counterclockwise ($-$) in the $xy$-plane and hence directed along the $z$-axis is given by $\ket{\bm\upsigma^{+ \pm}_{z}} = (1,\pm i,0)\ket{\bm\upsigma^+}$, where the axis of rotation and the rotation direction are marked by the subscript and the second sign in the superscript, respectively, and $\exp(-i \omega t)$ time dependence is assumed. The emission from the rotating chiral dipoles retains its pure handedness, but surprisingly it is unidirectional with the direction depending on the rotation (spin vector) direction (Figs.~\ref{rotatingchiral}(c--f)). If the handedness of a chiral dipole is changed to the opposite while keeping its spin the same, the radiation direction is changed to the opposite. 
 In all cases, the emission is predominantly perpendicular to the rotation plane. Mnemonically, the handedness of the chiral dipole, its rotation direction and the emission direction obey the hand rule: with the left or right hand chosen to match the handedness of the chiral dipole and the fingers are bent to indicate the direction of the rotation, the thumb will show the direction of the radiation. Mathematically, this can be expressed as $\mathbf{I} = \pm \mathbf{s}$, where $\mathbf{I}$ marks the direction of radiation, $\mathbf{s}$ is the spin vector of the rotating chiral dipole, related to the rotation direction by the right-hand rule, and the choice of $+$ or $-$ is defined by the handedness of the dipole. Consequently, for right-handed rotating chiral dipoles the emission is along their spin, and for left-handed counterparts opposite. In the case of chiral dipoles with 'elliptical' rotation (the combination of elliptically polarised magnetic and electric dipoles), the ratio of the forward and backward scattered intensities can be controlled by the phase difference between the $x$- and $y$- chiral dipole components, with equal intensities achieved in the case of a linear chiral dipole when the components are in phase (Fig.~\ref{rotatingchiral}g).    
 
 The introduced concept of rotating chiral dipoles can be linked to other common electromagnetic dipolar sources. Generally, a pair of electric and magnetic dipoles or a pair of chiral dipoles of different handedness forms a complete basis for representing any dipolar state along a given direction: $\ket{\mathbf{d}} = \mathbf{p}\ket{\mathbf{p}} + \mathbf{m}\ket{\mathbf{m}} = \bm\upsigma^+\ket{\bm\upsigma^+} + \bm\upsigma^-\ket{\bm\upsigma^-}$, where $\mathbf{p} = (p_x,p_y,p_z)$ and similarly for other amplitude vectors involved.

 A basis of four chiral dipoles or four rotating chiral dipoles is needed to produce an arbitrary in-plane dipolar state, while six states from either of these two bases form a complete set for the description of any dipolar state in three dimensions (see Supplementary Section~1 for details). For example, the rotating electric (magnetic) dipoles can be represented as two rotating out-of-phase (in-phase) chiral dipoles with the same spin and opposite handedness, while a Huygens' dipole, unidirectionally emitting a linearly polarised light (Supplementary Fig.~1b), can be viewed as two rotating chiral dipoles with opposite spins and handedness.
\subsection{Rotating chiral dipoles in plasmonic nanohelicoids}

\begin{figure*}[!ht]
    \begin{center}
        \includegraphics[width=1.\textwidth]{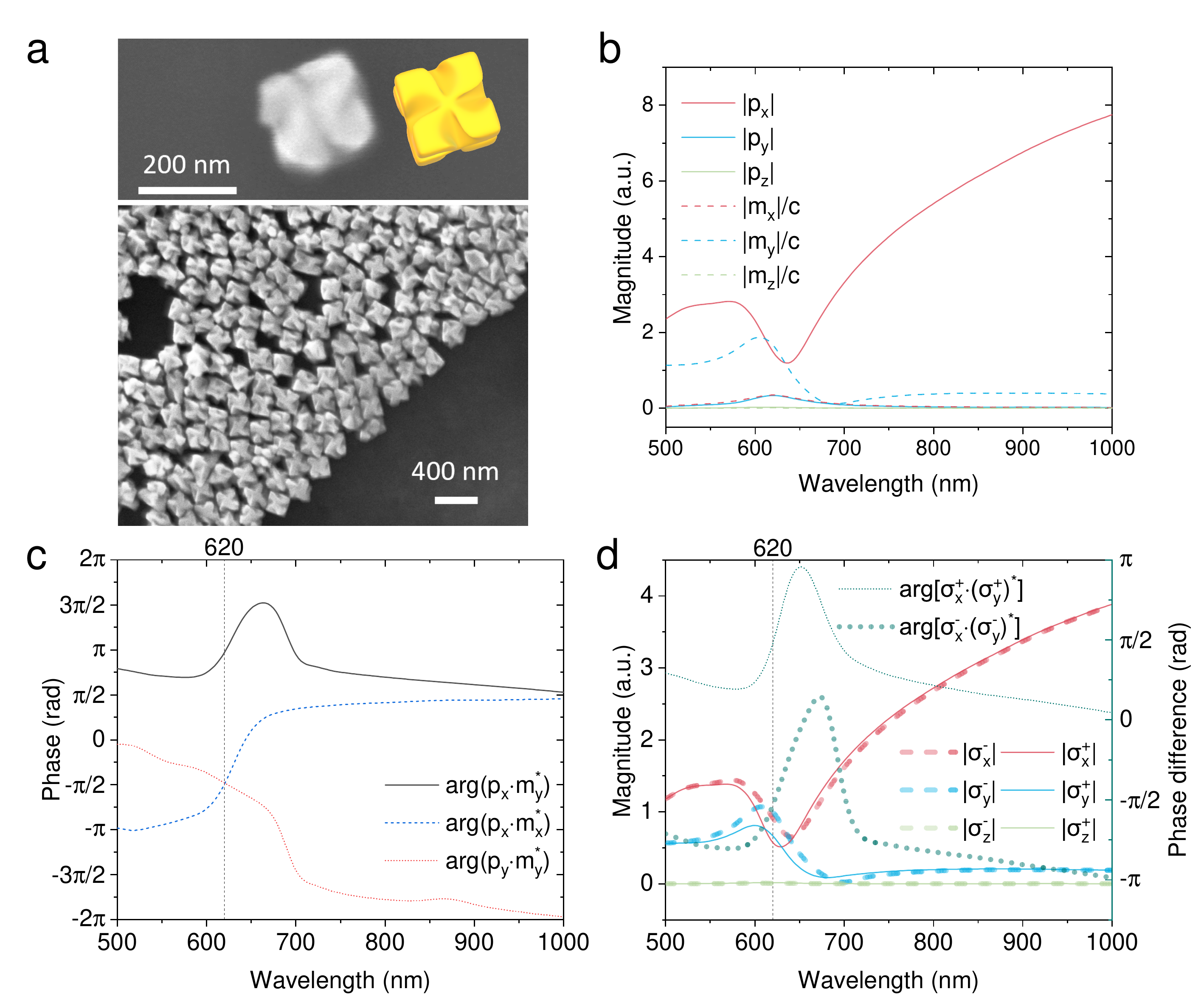}
        \caption{\textbf{Optical properties of plasmonic L-nanohelicoids.} \textbf{a}~Scanning electron microscopy images (top view) of an individual gold L-nanohelicoid (schematics is shown in yellow) and a L-nanohelicoid assembly on a substrate showing the uniformity of the fabricated nanostructures; the average size of the nanohelicoids is around 180~nm. \textbf{b}--\textbf{d}~Numerically calculated (\textbf{b})~magnitudes and (\textbf{c})~relative phases of the electric and magnetic dipoles excited in a 180~nm gold L-nanohelicoid in a \ch{SiO_2} matrix with a $x$-polarised plane wave propagating in the $-z$ direction, and (\textbf{d}) corresponding magnitudes and phases of the excited chiral dipoles.}\label{linear}
    \end{center}
\end{figure*}

Among various realisations of chiral nanoparticles \cite{vestler2018enhanced,Wang2022chiralsensing,han2023neural,ahn2019bioinspired,krasavin2005gammadion}, plasmonic nanohelicoids (Fig.~\ref{linear}a) provide strong resonant chiroptical effects \cite{kim2022enantioselective,lee2018amino,bainova2023plasmon,gao2021selective,chen2023inversion,Wang_ACSAMI_2024}. They can be either right-handed (D-nanohelicoids) or left-handed (L-nanohelicoids)~\cite{,chen2023inversion,Wang_ACSAMI_2024}, which is defined by the handedness of glutathione used for their fabrication (see Methods for the fabrication details). Broad spectral tuneability of their resonant response is favourable for designing electric and magnetic dipolar resonances with the required relative amplitude and phase, which makes them particularly suitable for the realisation of chiral dipoles. 

At the wavelength of the magnetic dipolar resonance, the optical properties of L-nanohelicoids of a 180 nm size in a \ch{SiO_2} matrix considered here are governed by the electric and magnetic dipolar responses (see Supplementary Information Section~2 for the details of the multipole decomposition, showing only a minor contribution from an electric quadrupole and no contribution from its magnetic counterpart (Supplementary Fig.~7b)). The contribution from the magnetic dipole is strongest if the handedness of the excitation light and the nanoparticle is the same, and suppressed in the opposite case.

Particularly, under illumination with a $x$-polarised plane wave propagating in the $z$-direction, the dominant contributions come from an electric dipole with a magnitude $|p_x|$ parallel to the direction of the incident electric field and a magnetic dipole with a magnitude $|m_y/c|$ perpendicular to it (Fig.~\ref{linear}b). Near the magnetic resonance at a wavelength of around 620~nm, their magnitudes are close to each other. Simultaneously, the phase between them is $\pi$, i.e. they are out of phase (Fig.~\ref{linear}c). These conditions correspond to the excitation of a Huygens dipole at this wavelength. There are no electric or magnetic dipole components excited along the propagation direction of the incident wave ($z$-direction). At the same time, and it is very important, there exists also a magnetic dipole with a magnitude $|m_x|$ parallel to the dominant $|p_x|$ electric counterpart and an electric dipole with a magnitude $|p_y|$ parallel to the dominant $|m_y|$ magnetic counterpart. Furthermore, the phase shifts between $p_x$ and $m_x$ as well as between $p_y$ and $m_y$ at the resonant wavelength of 620~nm are $-\pi/2$ (see dashed blue and red curves in Fig.~\ref{linear}c, respectively), which signifies the formation of two $\ket{\bm\upsigma^-}$ chiral dipoles along the $x$- and $y$- directions and their dominance over their $\ket{\bm\upsigma^+}$ counterparts. Such predominance is the origin of the overall chiroptical response related to the geometrical chirality of the nanohelicoid. This can be seen directly by decomposition of the excited dipolar state into the chiral dipole basis $\bm\upsigma^{\pm} = (\mathbf{m}/c \mp i \mathbf{p})/2$ 

(see Supplementary Information Section~3 for details) as a difference in the excitation of right- and left-handed chiral dipoles. Comparing red thick dashed line for $|\sigma^-_x|$ with red thin solid line for $|\sigma^+_x|$, and blue thick dashed line for $|\sigma^-_y|$ with blue thin solid line for $|\sigma^+_y|$ in Fig.~\ref{linear}d one can see that
\begin{equation}\label{unequal}
\begin{split}
|\sigma^-_x| \approx |\sigma^-_y| > |\sigma^+_x| \approx |\sigma^+_y|
\end{split}
\end{equation}
which marks the dominant excitation of $\ket{\bm\upsigma^-}$ and the appearance of chirooptical effects. Furthermore, in the rotating chiral dipole basis, the relative phases between $\ket{\bm\upsigma^\pm}$ components along the 
$x$- and $y$- axes $\text{arg}[\sigma^+_x \cdot (\sigma^+_y)^\ast] = \pi/2$ and $\text{arg}[\sigma^-_x \cdot (\sigma^-_y)^\ast] = -\pi/2$ signify the excitation of both $\ket{\bm\upsigma^{+-}_{z}}$ and $\ket{\bm\upsigma^{-+}_{z}}$ rotating chiral dipoles under linear polarised excitation, but with different magnitudes defined by the object chirality.

\begin{figure*}[!ht]
    \begin{center}
        \includegraphics[width=1.\textwidth]{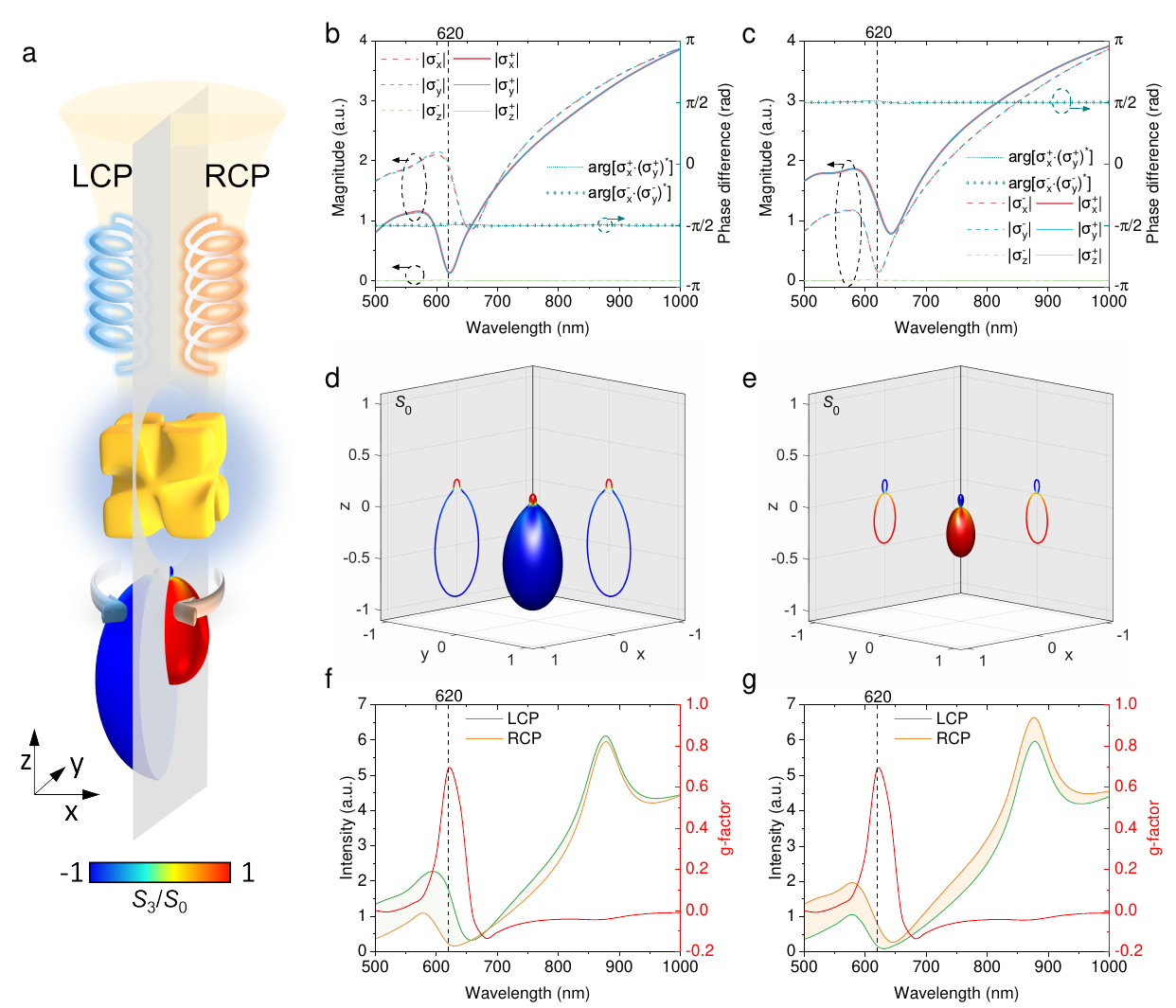}
        \caption{\textbf{Enantio-sensitive excitation of rotating chiral dipoles by circularly polarised light.} \textbf{a}~The schematics of the excitation of rotating chiral dipoles in a L-nanohelicoid by circularly polarised light with the corresponding simulated far-field scattering intensity and polarisation diagrams. \textbf{b},~\textbf{c}~Numerically calculated magnitudes and relative phases of the chiral dipoles excited in a 180~nm gold L-nanohelicoid in \ch{SiO_2} with (\textbf{b}) a LCP and (\textbf{c}) a RCP plane wave propagating in the $-z$ direction. \textbf{d},~\textbf{e}~Far-field scattering intensity and polarisation diagrams calculated at the wavelength of 620~nm in cases (b) and (c), respectively. The colour indicates the polarisation state of the scattered light calculated from Stokes parameters $S_3/S_0$ (1: RCP; -1: LCP). \textbf{f},~\textbf{g}~Integrated intensities for RCP and LCP polarisations of the scattered light in cases (b) and (c), respectively.}\label{ht200}
    \end{center}
\end{figure*}

Under circularly polarised illumination, a second set of electric and magnetic dipoles (orthogonal counterparts of that presented in Fig.~\ref{linear}b) will be excited with a phase delay determined by the handedness of the incident light polarisation. Generally, this will result in a sum of elliptically polarised electric and magnetic dipoles with certain elliptical trajectories and phase delays. However, at a resonant wavelength of 620~nm light, 
$\ket{\bm\upsigma^{-+}_{z}}$ is excited in the case of LCP illumination and $\ket{\bm\upsigma^{+-}_{z}}$ in the case of RCP, 
because of balanced excitation of electric and magnetic dipoles in nanohelicoids (Fig.~\ref{linear}b): for LCP (RCP) illumination $\ket{\bm\upsigma^-_x}$ and $\ket{\bm\upsigma^-_y}$ ($\ket{\bm\upsigma^+_x}$ and $\ket{\bm\upsigma^+_y}$) are predominantly excited, importantly with the same magnitudes and a $-\pi/2$ ($\pi/2$) phase difference, resulting in the rotation direction coinciding with the handedness of the incident field (Fig.~\ref{ht200}b,c).

Thus, the direction of the spin of the excited rotating chiral dipoles is determined by the handedness and the propagation direction of the illuminating light. This can be further proved analytically (see Supplementary Information Section~4 for details).
At the same time, the magnitudes of the excited $\ket{\bm\upsigma^-}$ and $\ket{\bm\upsigma^+}$ chiral dipoles (as can be seen from Eq.~\ref{unequal} for $x$-polarised excitation), and therefore the magnitudes of the resulting rotating chiral dipoles are unequal for the LCP and RCP, which signifies the interaction of the chirality of the illuminating light with the chirality of the nanohelicoid, i.e. the enantio-sensitivity and chiral nature of the scattering process.

Nanosphere and nanocube particles of the same size show predominantly electric dipolar and electric quadrupolar optical responses, with only a small contribution of magnetic dipoles (cf. Supplementary Figs.~2, 3 and 7), which hinders the excitation of pure rotating chiral dipole states. A certain directionality of the scattering in the case of the nanocubes is the result of favourable interference of electric dipolar and electric quadrupolar scattering \cite{liu2018generalized}. However, as these nanoparticles are achiral, they interact with light of different circular polarisations in the same fashion.

From the radiation patterns of the ideal rotating chiral dipoles (Fig.~\ref{rotatingchiral}c--f), one can see that the ratio between the magnitudes of the excited rotating chiral dipoles defines the balance between the forward scattering with the same polarisation as the incident one and backward scattering with the opposite handedness. As this ratio is wavelength-dependent (Fig.~\ref{ht200}b,c), gradual spectral evolution of the scattering pattern is observed (Fig.~\ref{rotatingchiral}g). Near the magnetic resonance at a wavelength of 620~nm (Figs.~\ref{ht200}b,c), one dominant rotating chiral dipole is excited, while the amplitude of a (secondary) rotating chiral dipole with the opposite handedness is minimised. The 180-nm nanohelicoid size chosen in the study offers the highest contrast between the opposite handednesses (Supplementary Fig.~8).  
The resulting emission directionality is similar to the ideal case shown in Fig.~\ref{rotatingchiral}c,f,  revealing vastly predominant forward scattering with the same handedness as the illuminating light (Fig.~\ref{ht200}d,e). 
At the same time, since a nanohelicoid is a chiral object, it interacts differently with the circularly polarised light of different handedness. Particularly, for a given chirality of the nanohelicoid, RCP and LCP light is forward-scattered with unequal efficiency. When the direction of the rotation of the electric field of the incident wave coincides with the direction of the nanohelicoid twist (spin of the excited rotating chiral dipole and chirality of the nanohelicoid are aligned), forward scattering is stronger. Both effects, the dominant scattering of the light of the same handedness as the incident light and the role of nanohelicoid chirality (cf. LCP in Fig.~\ref{ht200}f and RCP in Fig.~\ref{ht200}g) can be observed for the total emitted intensities (Fig.~\ref{ht200}f,g). It is also interesting to note that the chiroptical response expressed as a conventional g-factor
\begin{equation}\label{g-factor}
g_{\text{scat}} = 2 \frac{C_{\text{scat}}^{\text{LCP}}-C_{\text{scat}}^{\text{RCP}}}{C_{\text{scat}}^{\text{LCP}}+C_{\text{scat}}^{\text{RCP}}},
\end{equation}
where $C_{\text{scat}}^{\text{LCP}}$ and $C_{\text{scat}}^{\text{RCP}}$ are the scattering cross-sections for LCP and RCP light, respectively, reaches its maximum at the 620~nm wavelength at which pure chiral dipoles are excited (same can be observed for extinction g-factor, see Supplementary Fig.~9). This gives a confirmation that the efficiency of the excitation of pure rotating chiral dipolar states is influenced by the presence of structural chirality. The scattered fields can be also affected by the excitation of higher-order multipoles, e.g. the peak just above 860~nm corresponds to the electric quadrupolar resonance (Supplementary Fig.~10). Making an impact on the polarisation state of the scattered light, they can smear-out the contribution from a pure chiral dipole, resulting in the 
decrease of the g-factor. Note that comparing Supplementary Figs.~7a and c, one can see that the contributions of the excited magnetic dipole and excited electric dipole are the same when the handedness of the illuminating light matches the handedness of the nanohelicoid (at the same time, for the opposite handedness, the contribution of a magnetic dipole is reduced). Consequently, a well-defined rotating chiral dipole is excited, not only providing a larger scattering intensity, but also a better directionality (cf. Figs.~\ref{ht200}d and \ref{ht200}e) and a better LCP/RCP contrast (cf. Figs.~\ref{ht200}f and \ref{ht200}g).

\subsection{Experimental demonstration}

\begin{figure*}[!ht]
    \begin{center}
        \includegraphics[width=1.\textwidth]{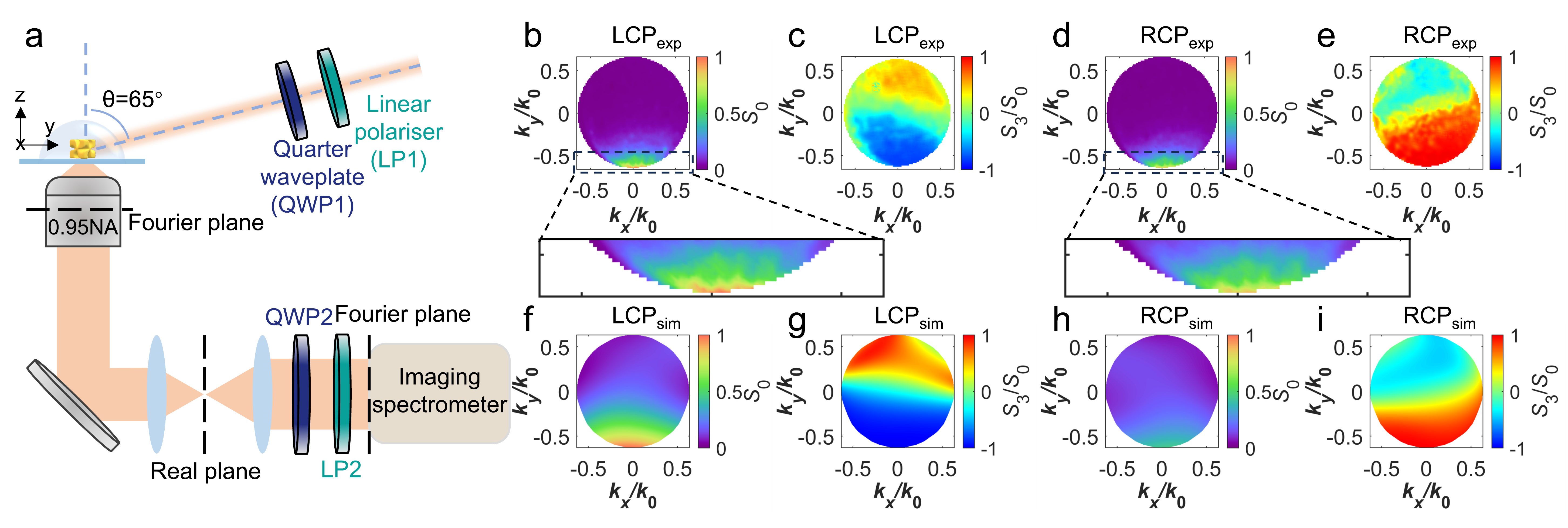}
        \caption{\textbf{Experimental observation of chiral scattering from single L-nanohelicoid.} \textbf{a}~Sketch of the experimental setup. \textbf{b}--\textbf{e}~Experimental and \textbf{f}--\textbf{i} numerical Fourier-space maps of normalised $S_0$ (\textbf{b},\textbf{f},\textbf{d},\textbf{h}) showing the intensity of the scattered light and $S_3/S_0$ (\textbf{c},\textbf{g},\textbf{e},\textbf{i}) showing the polarisation for (\textbf{b},\textbf{c},\textbf{f},\textbf{g})~LCP and (\textbf{d},\textbf{h},\textbf{e},\textbf{i})~RCP illumination.} \label{experiment}
    \end{center} 
\end{figure*}

For experimental demonstration, gold nanohelicoids with a size of 180~nm were prepared by an amino-guided growth method \cite{lee2018amino} (see Methods). They were deposited on a silica coverslip by spin coating, during which the nanohelicoids were well separated, allowing to make measurements on individual particles (Supplementary Fig.~11). Then, they were further covered with a \ch{SiO_2} layer to provide a uniform surrounding environment. Angle-resolved scattering from a single nanohelicoid for both circular polarisations were measured using a Fourier plane imaging technique with 610~nm illumination (Fig.~\ref{experiment}a) (see Methods for the details of the experimental set-up). The illumination wavelength was chosen to match the maximum of the experimentally measured circular dichroism of the nanohelicoids in \ch{SiO2} (Supplementary Fig.~12). The maps of $S_{0}$ and $S_{3}$ plotted for both LCP and RCP polarisations of the incident beam and representing the intensity and polarisation of the scattered light, show predominant forward scattering with a high-quality polarisation state with the same handedness as that of the incident light (Fig.~\ref{experiment}b--e), confirming the numerical predictions and their analysis based on the excitation of rotating chiral dipoles (Fig.~\ref{experiment}f--i). Importantly, compared to the scattering of RCP incident light, the LCP illumination results in stronger forward scattering, signifying the chiral nature of the scattering. The relative difference of the scattered intensities of LCP and RCP light in the zoom-in area $2(I_{\text{LCP}}-I_{\text{RCP}})/(I_{\text{LCP}}+I_{\text{RCP}})$ corresponds to $\sim$~15\%, which is lower than the theoretical prediction of $\sim$50\%, due to the deviations of the helicoid shapes from the ideal modelled one, resulting in the breaking of the balance between matched amplitudes of the excited electric and magnetic dipoles at the resonant wavelength (see Fig.~\ref{linear}b for the case of the $x$-polarised component).   

 In the single-particle scattering measurements, the orientation of the nanohelicoid with regard to the laser beam is generally unknown. To take this into account, numerical simulations for characteristic spatial orientations of the nanohelicoid with respect to the direction of the illumination light were performed (Supplementary Fig.~4). It was found that although there are some variations in the far-field scattering patterns due to a non-ideal balance of the magnitudes and phases of the excited electric and magnetic dipoles, the handedness of the forward scattered light and the ratio between scattered intensities of LCP and RCP light remain the same, which explains the robustness of the experimental results.

\section{Discussion}
We have introduced a concept of rotating chiral dipoles for achieving directional enantio-sensitive scattering from chiral nanoparticles. The scattering direction, handedness of scattered light and chiral dipole rotation are linked through the hand rule, which choice is determined by the chirality of the dipole. Such dipoles can be excited in nanoparticles by circularly polarised light with both handedness of the chiral dipole and its rotation spin determined by the spin of the illuminating light. The observations of the circularly polarised light scattering from a single plasmonic chiral nanoparticle reveal the excitation of pure rotating chiral dipole states, leading to vastly predominant forward scattering with the same handedness. The experimentally observed scattering is enantio-sensitive, demonstrating the chiral character of the optical interaction. Particularly, the scattering is enhanced if the chirality of the illuminating light and the nanoparticle is the same. Due to the amplitude and phase differences of the electric and magnetic dipoles of the nanohelicoid compared to those required for an ideal rotating chiral dipole, the scattering of light of opposite handedness is suppressed but present in the same directional manner.    

The obtained results can be useful for enantio-sensitive optomechanical manipulation (see Supplementary Section~7 for discussions of enantio-sensitive optical forces), chiral sensing and quantum optical technology.

\section{Methods}
\subsection{Fabrication of plasmonic nanohelicoids} 
The L-helicoid nanoparticles were grown from octahedral gold seeds using a method described in Ref.~\cite{lee2018amino}. The growth solution included 29.04~mL DI water, 6.4~mL 0.1~M CTAB and 640~$\upmu$L 10~mM  \ch{HAuCl4}. Then, a 3.6~mL 0.1~M ascorbic acid solution was injected into the growth solution as a reducing agent. Then, the chiral growth was triggered by adding 16~$\upmu$L 5~$\upmu$M of L-glutathione to a 320~$\upmu$L solution of the octahedral seeds. After 2~h in a water bath at $30^\circ$~C, the particles were washed and collected by centrifuging 3 times and kept in a 1~mM CTAB solution for further use.

The helicoid nanoparticles embedded in \ch{SiO2} were prepared in two steps. First, helicoid nanoparticles were dispersed on a plasma-cleaned glass by spin coating. Then, a \ch{SiO2} layer was deposited on the top of helicoid nanoparticles by a sol-gel spin coating method \cite{hinczewski2005optical}. Briefly, a \ch{SiO2} sol-gel solution was prepared mixing 223~$\upmu$L of TEOS, 992~mg of isopropanol, 278~$\upmu$L of DI water and 12.46 $\upmu$L of a 37\% HCl solution. The mixture was kept stirring for 80~min in a water bath at $70^\circ$~C before the spin coating.
\subsection{Finite element method numerical simulations}
Scattering of light from a single helicoidal nanoparticle 
%with a side length of 180~nm in \ch{SiO2} 
was simulated using the finite element method (COMSOL Multiphysics software) in a scattered field formulation. The nanohelicoid was positioned with its major axes aligned with the simulation reference frame and the wave vector of the illumination light wave was set in $-z$ direction (see Fig.~3 for the co-ordinate axis orientation). The dielectric permittivity of gold, water and \ch{SiO2} were taken from the tabulated experimental data \cite{PhysRevB.6.4370, hale1973optical,malitson1965interspecimen}. The simulation domain was surrounded by a perfectly matching layer to assure the absence of back-reflection of the scattered waves.

To calculate Stokes parameters, far-field components $E_{\theta}$ and $E_{\phi}$ of the radiated or scattered electric field were analysed. Stokes parameters $S_0$ and $S_3$ describing the total intensity and polarisation state of the field were obtained in a usual way as
$S_0= |E_{\theta}|^2 + |E_{\phi}|^2$ and
$S_3=-2\Im(E_{\theta} \cdot E_{\phi}^*)$.
\subsection{Fourier microscopy measurements}
 The sample was illuminated by a collimated white light from a supercontinuum laser (NKT Photonics SuperK-EVO-HP), filtered at the wavelength of interest, and circularly polarised using a broadband linear polariser and a quarter waveplate. A hemispherical lens together with index-matching oil was used to obtain the index-matching illumination conditions. The angle of incidence of the beam on the nanohelicoid was 65$^\circ$. Such an almost grazing angle allows the observation of both forward and backward scattering in the implemented Fourier setup. The scattered signal from the nanohelicoid was collected by a $100 \times$ microscope objective with NA$=0.95$, allowing collection within a large solid angle. The Fourier plane (back focal plane of the detection objective) was then imaged onto a CCD camera using a set of relay lenses. The polarisation of the scattered signal was analysed using a quarter waveplate and a linear polariser, allowing the reconstruction of the Stokes parameters.

\bibliographystyle{naturemag}
\bibliography{bib}

\begin{thebibliography}{10}
\expandafter\ifx\csname url\endcsname\relax
  \def\url#1{\texttt{#1}}\fi
\expandafter\ifx\csname urlprefix\endcsname\relax\def\urlprefix{URL }\fi
\providecommand{\bibinfo}[2]{#2}
\providecommand{\eprint}[2][]{\url{#2}}

\bibitem{goerlitzer2021beginner}
\bibinfo{author}{Goerlitzer, E.~S.}, \bibinfo{author}{Puri, A.~S.}, \bibinfo{author}{Moses, J.~J.}, \bibinfo{author}{Poulikakos, L.~V.} \& \bibinfo{author}{Vogel, N.}
\newblock \bibinfo{title}{The beginner's guide to chiral plasmonics: Mostly harmless theory and the design of large-area substrates}.
\newblock \emph{\bibinfo{journal}{Adv. Opt. Mater.}} \bibinfo{pages}{2100378} (\bibinfo{year}{2021}).

\bibitem{ke2023vacuum}
\bibinfo{author}{Ke, Y.}, \bibinfo{author}{Song, Z.} \& \bibinfo{author}{Jiang, Q.-D.}
\newblock \bibinfo{title}{Vacuum-induced symmetry breaking of chiral enantiomer formation in chemical reactions}.
\newblock \emph{\bibinfo{journal}{Phys. Rev. Lett.}} \textbf{\bibinfo{volume}{131}}, \bibinfo{pages}{223601} (\bibinfo{year}{2023}).

\bibitem{luo2023large}
\bibinfo{author}{Luo, J.} \emph{et~al.}
\newblock \bibinfo{title}{Large effective magnetic fields from chiral phonons in rare-earth halides}.
\newblock \emph{\bibinfo{journal}{Science}} \textbf{\bibinfo{volume}{382}}, \bibinfo{pages}{698--702} (\bibinfo{year}{2023}).

\bibitem{wang2017circular}
\bibinfo{author}{Wang, X.} \& \bibinfo{author}{Tang, Z.}
\newblock \bibinfo{title}{Circular dichroism studies on plasmonic nanostructures}.
\newblock \emph{\bibinfo{journal}{Small}} \textbf{\bibinfo{volume}{13}}, \bibinfo{pages}{1601115} (\bibinfo{year}{2017}).

\bibitem{bliokh2016spin}
\bibinfo{author}{Bliokh, K.~Y.} \emph{et~al.}
\newblock \bibinfo{title}{Spin-hall effect and circular birefringence of a uniaxial crystal plate}.
\newblock \emph{\bibinfo{journal}{Optica}} \textbf{\bibinfo{volume}{3}}, \bibinfo{pages}{1039--1047} (\bibinfo{year}{2016}).

\bibitem{eismann2018exciting}
\bibinfo{author}{Eismann, J.~S.}, \bibinfo{author}{Neugebauer, M.} \& \bibinfo{author}{Banzer, P.}
\newblock \bibinfo{title}{Exciting a chiral dipole moment in an achiral nanostructure}.
\newblock \emph{\bibinfo{journal}{Optica}} \textbf{\bibinfo{volume}{5}}, \bibinfo{pages}{954--959} (\bibinfo{year}{2018}).

\bibitem{picardi2022integrated}
\bibinfo{author}{Picardi, M.~F.} \emph{et~al.}
\newblock \bibinfo{title}{Integrated janus dipole source for selective coupling to silicon waveguide networks}.
\newblock \emph{\bibinfo{journal}{Appl. Phys. Rev.}} \textbf{\bibinfo{volume}{9}} (\bibinfo{year}{2022}).

\bibitem{liu2018generalized}
\bibinfo{author}{Liu, W.} \& \bibinfo{author}{Kivshar, Y.~S.}
\newblock \bibinfo{title}{Generalized kerker effects in nanophotonics and meta-optics}.
\newblock \emph{\bibinfo{journal}{Opt. Express}} \textbf{\bibinfo{volume}{26}}, \bibinfo{pages}{13085--13105} (\bibinfo{year}{2018}).

\bibitem{liu2012broadband}
\bibinfo{author}{Liu, W.}, \bibinfo{author}{Miroshnichenko, A.~E.}, \bibinfo{author}{Neshev, D.~N.} \& \bibinfo{author}{Kivshar, Y.~S.}
\newblock \bibinfo{title}{Broadband unidirectional scattering by magneto-electric core--shell nanoparticles}.
\newblock \emph{\bibinfo{journal}{ACS nano}} \textbf{\bibinfo{volume}{6}}, \bibinfo{pages}{5489--5497} (\bibinfo{year}{2012}).

\bibitem{elancheliyan2020tailored}
\bibinfo{author}{Elancheliyan, R.} \emph{et~al.}
\newblock \bibinfo{title}{Tailored self-assembled nanocolloidal huygens scatterers in the visible}.
\newblock \emph{\bibinfo{journal}{Nanoscale}} \textbf{\bibinfo{volume}{12}}, \bibinfo{pages}{24177--24187} (\bibinfo{year}{2020}).

\bibitem{kuznetsov2016optically}
\bibinfo{author}{Kuznetsov, A.~I.}, \bibinfo{author}{Miroshnichenko, A.~E.}, \bibinfo{author}{Brongersma, M.~L.}, \bibinfo{author}{Kivshar, Y.~S.} \& \bibinfo{author}{Luk’yanchuk, B.}
\newblock \bibinfo{title}{Optically resonant dielectric nanostructures}.
\newblock \emph{\bibinfo{journal}{Science}} \textbf{\bibinfo{volume}{354}}, \bibinfo{pages}{aag2472} (\bibinfo{year}{2016}).

\bibitem{picardi2019experimental}
\bibinfo{author}{Picardi, M.~F.} \emph{et~al.}
\newblock \bibinfo{title}{Experimental demonstration of linear and spinning janus dipoles for polarisation-and wavelength-selective near-field coupling}.
\newblock \emph{\bibinfo{journal}{Light Sci. Appl.}} \textbf{\bibinfo{volume}{8}}, \bibinfo{pages}{52} (\bibinfo{year}{2019}).

\bibitem{ballantine2020optical}
\bibinfo{author}{Ballantine, K.} \& \bibinfo{author}{Ruostekoski, J.}
\newblock \bibinfo{title}{Optical magnetism and huygens’ surfaces in arrays of atoms induced by cooperative responses}.
\newblock \emph{\bibinfo{journal}{Phys. Rev. Lett.}} \textbf{\bibinfo{volume}{125}}, \bibinfo{pages}{143604} (\bibinfo{year}{2020}).

\bibitem{kaelberer2010toroidal}
\bibinfo{author}{Kaelberer, T.}, \bibinfo{author}{Fedotov, V.}, \bibinfo{author}{Papasimakis, N.}, \bibinfo{author}{Tsai, D.} \& \bibinfo{author}{Zheludev, N.}
\newblock \bibinfo{title}{Toroidal dipolar response in a metamaterial}.
\newblock \emph{\bibinfo{journal}{Science}} \textbf{\bibinfo{volume}{330}}, \bibinfo{pages}{1510--1512} (\bibinfo{year}{2010}).

\bibitem{meng2022measuring}
\bibinfo{author}{Meng, F.} \emph{et~al.}
\newblock \bibinfo{title}{Measuring the magnetic topological spin structure of light using an anapole probe}.
\newblock \emph{\bibinfo{journal}{Light sci. appl.}} \textbf{\bibinfo{volume}{11}}, \bibinfo{pages}{287} (\bibinfo{year}{2022}).

\bibitem{miroshnichenko2015nonradiating}
\bibinfo{author}{Miroshnichenko, A.~E.} \emph{et~al.}
\newblock \bibinfo{title}{Nonradiating anapole modes in dielectric nanoparticles}.
\newblock \emph{\bibinfo{journal}{Nat. Commun.}} \textbf{\bibinfo{volume}{6}}, \bibinfo{pages}{8069} (\bibinfo{year}{2015}).

\bibitem{vestler2018enhanced}
\bibinfo{author}{Vestler, D.} \emph{et~al.}
\newblock \bibinfo{title}{Circular dichroism enhancement in plasmonic nanorod metamaterials}.
\newblock \emph{\bibinfo{journal}{Opt. Express}} \textbf{\bibinfo{volume}{26}}, \bibinfo{pages}{17841--17848} (\bibinfo{year}{2018}).

\bibitem{Wang2022chiralsensing}
\bibinfo{author}{Wang, P.} \emph{et~al.}
\newblock \bibinfo{title}{Molecular plasmonics with metamaterials}.
\newblock \emph{\bibinfo{journal}{Chem. Rev.}} \textbf{\bibinfo{volume}{122}}, \bibinfo{pages}{15031--15081} (\bibinfo{year}{2022}).

\bibitem{han2023neural}
\bibinfo{author}{Han, J.~H.} \emph{et~al.}
\newblock \bibinfo{title}{Neural-network-enabled design of a chiral plasmonic nanodimer for target-specific chirality sensing}.
\newblock \emph{\bibinfo{journal}{ACS Nano}} \textbf{\bibinfo{volume}{17}}, \bibinfo{pages}{2306--2317} (\bibinfo{year}{2023}).

\bibitem{ahn2019bioinspired}
\bibinfo{author}{Ahn, H.-Y.} \emph{et~al.}
\newblock \bibinfo{title}{Bioinspired toolkit based on intermolecular encoder toward evolutionary {4D} chiral plasmonic materials}.
\newblock \emph{\bibinfo{journal}{Acc. Chem. Res.}} \textbf{\bibinfo{volume}{52}}, \bibinfo{pages}{2768--2783} (\bibinfo{year}{2019}).

\bibitem{krasavin2005gammadion}
\bibinfo{author}{Krasavin, A.} \emph{et~al.}
\newblock \bibinfo{title}{Polarization conversion and ``focusing{''} of light propagating through a small chiral hole in a metallic screen}.
\newblock \emph{\bibinfo{journal}{Appl. Phys. Lett.}} \textbf{\bibinfo{volume}{86}}, \bibinfo{pages}{201105} (\bibinfo{year}{2005}).

\bibitem{kim2022enantioselective}
\bibinfo{author}{Kim, R.~M.} \emph{et~al.}
\newblock \bibinfo{title}{Enantioselective sensing by collective circular dichroism}.
\newblock \emph{\bibinfo{journal}{Nature}} \textbf{\bibinfo{volume}{612}}, \bibinfo{pages}{470--476} (\bibinfo{year}{2022}).

\bibitem{lee2018amino}
\bibinfo{author}{Lee, H.-E.} \emph{et~al.}
\newblock \bibinfo{title}{Amino-acid-and peptide-directed synthesis of chiral plasmonic gold nanoparticles}.
\newblock \emph{\bibinfo{journal}{Nature}} \textbf{\bibinfo{volume}{556}}, \bibinfo{pages}{360--365} (\bibinfo{year}{2018}).

\bibitem{bainova2023plasmon}
\bibinfo{author}{Bainova, P.} \emph{et~al.}
\newblock \bibinfo{title}{Plasmon-assisted chemistry using chiral gold helicoids: Toward asymmetric organic catalysis}.
\newblock \emph{\bibinfo{journal}{ACS Catal.}} \textbf{\bibinfo{volume}{13}}, \bibinfo{pages}{12859--12867} (\bibinfo{year}{2023}).

\bibitem{gao2021selective}
\bibinfo{author}{Gao, H.}, \bibinfo{author}{Chen, P.-G.}, \bibinfo{author}{Lo, T.~W.}, \bibinfo{author}{Jin, W.} \& \bibinfo{author}{Lei, D.}
\newblock \bibinfo{title}{Selective excitation of polarization-steered chiral photoluminescence in single plasmonic nanohelicoids}.
\newblock \emph{\bibinfo{journal}{Adv. Funct. Mater.}} \textbf{\bibinfo{volume}{31}}, \bibinfo{pages}{2101502} (\bibinfo{year}{2021}).

\bibitem{chen2023inversion}
\bibinfo{author}{Chen, Y.} \emph{et~al.}
\newblock \bibinfo{title}{Inversion of the chiroptical responses of chiral gold nanoparticles with a gold film}.
\newblock \emph{\bibinfo{journal}{ACS nano}} \textbf{\bibinfo{volume}{18}}, \bibinfo{pages}{383--394} (\bibinfo{year}{2024}).

\bibitem{Wang_ACSAMI_2024}
\bibinfo{author}{Wang, L.} \emph{et~al.}
\newblock \bibinfo{title}{Circular differential photocurrent mapping of hot electron response from individual plasmonic nanohelicoids}.
\newblock \emph{\bibinfo{journal}{ACS Appl. Mater. Interfaces.}}  (\bibinfo{year}{2024}).
\newblock \eprint{https://doi.org/10.1021/acsami.4c03457}.

\bibitem{hinczewski2005optical}
\bibinfo{author}{Hinczewski, D.~S.}, \bibinfo{author}{Hinczewski, M.}, \bibinfo{author}{Tepehan, F.} \& \bibinfo{author}{Tepehan, G.}
\newblock \bibinfo{title}{Optical filters from {SiO$_2$} and {TiO$_2$} multi-layers using sol--gel spin coating method}.
\newblock \emph{\bibinfo{journal}{Sol. Energy Mater.}} \textbf{\bibinfo{volume}{87}}, \bibinfo{pages}{181--196} (\bibinfo{year}{2005}).

\bibitem{PhysRevB.6.4370}
\bibinfo{author}{Johnson, P.~B.} \& \bibinfo{author}{Christy, R.~W.}
\newblock \bibinfo{title}{Optical constants of the noble metals}.
\newblock \emph{\bibinfo{journal}{Phys. Rev. B}} \textbf{\bibinfo{volume}{6}}, \bibinfo{pages}{4370--4379} (\bibinfo{year}{1972}).

\bibitem{hale1973optical}
\bibinfo{author}{Hale, G.~M.} \& \bibinfo{author}{Querry, M.~R.}
\newblock \bibinfo{title}{Optical constants of water in the 200-nm to 200-$\mu$m wavelength region}.
\newblock \emph{\bibinfo{journal}{Appl. Opt.}} \textbf{\bibinfo{volume}{12}}, \bibinfo{pages}{555--563} (\bibinfo{year}{1973}).

\bibitem{malitson1965interspecimen}
\bibinfo{author}{Malitson, I.~H.}
\newblock \bibinfo{title}{Interspecimen comparison of the refractive index of fused silica}.
\newblock \emph{\bibinfo{journal}{J. Opt. Soc. Am.}} \textbf{\bibinfo{volume}{55}}, \bibinfo{pages}{1205--1209} (\bibinfo{year}{1965}).

\end{thebibliography}
\section*{Acknowledgements}

This work was supported by the ERC iCOMM project (789340) and the UK EPSRC project (EP/W017075/1). Y.X. acknowledges support from China Scholarship Council. The authors are grateful to Prof.~Ki Tae Nam for providing a CAD geometry of the nanohelicoid (432 helicoid III) and to Dr.~Vittorio Aita for help with the experimental setup.

\section*{Author contributions statement}
Y.X., A.V.K and A.V.Z. developed the concept, Y.X. and A.V.K. performed the simulations, Y.X. fabricated nanoparticles, Y.X. and D.J.R. performed the measurements, all authors contributed to writing of the manuscript. 

\end{document}